\documentclass[article,5p]{elsarticle}

\pdfoutput=1
\makeatletter
\def\ps@pprintTitle{%
	\let\@oddhead\@empty
	\let\@evenhead\@empty
	\def\@oddfoot{}%
	\let\@evenfoot\@oddfoot}
\makeatother
\usepackage{lineno}
\usepackage{hyperref}
\usepackage{geometry}
\usepackage{physics}
\usepackage{amsmath}
\usepackage{amssymb}
\usepackage[T1]{fontenc}
\usepackage{etoolbox}
\usepackage[bottom]{footmisc}
\modulolinenumbers[5]

\title{Kaluza-Klein theories in the framework of Polymer Quantum Mechanics}

\author[1,2]{Giovanni Montani\corref{ead1}}
\ead{giovanni.montani@enea.it}
\author[2]{Sebastiano Segreto\corref{ead2}}
\ead{segreto.1821554@studenti.uniroma1.it}

\cortext[ead1]{\href{mailto:giovanni.montani@enea.it}{giovanni.montani@enea.it}}
\cortext[ead2]{\href{mailto:segreto.1821554@studenti.uniroma1.it}{segreto.1821554@studenti.uniroma1.it}}

\address[1]{ENEA, FSN-FUSPHY-TSM, R.C. Frascati, Via E. Fermi 45, 00044 Frascati, Italy}
\address[2]{Physics Department, “Sapienza” University of Rome, P.le Aldo Moro 5, 00185 (Roma), Italy}

\begin{document}
	
	\begin{frontmatter}
		
		\begin{abstract}
			
			We provide a re-analysis of the $5D$ Kaluza-Klein theory by implementing the polymer representation of the dynamics, both on a classical and a quantum level, in order to introduce in the model information about the existence of a cut-off scale. We start by showing that, in the framework of semi-classical quantum mechanics, the $5D$ Bianchi I model admits a solution in which three space directions expand isotropically, while the remaining one is static, offering in this way a very valuable scenario to implement a Kaluza-Klein paradigm, identifying in such a static dimension the compactified one. We then analyse the behaviour of geodesic motion in the context of the polymer representation, as referred to a $5D$ space-time with a static dimension. We demonstrate that such a revised formulation allows overcoming one of the puzzling questions of the standard Kaluza-Klein model corresponding to the limit of the charge to mass ratio for a particle, inapplicable to any fundamental one. Indeed, here, such a ratio can be naturally attributed to 
			particles predicted by the Standard Model and no internal contradiction of the theory arises on this level.
			
			Finally, we study the morphology of the field equation associated with a charged scalar particle, i.e. we analyse a Klein-Gordon equation, which fifth coordinate is viewed in the polymer representation. 
			Here we obtain the surprising result that, although the Kaluza-Klein tower has a 
			deformed structure characterized by irregular steps, the value predicted for the particle mass can be, in principle, set within the Standard Model mass distribution and hence, the problem of the Planckian value of such mass, typical of the standard formulation, is now overcome. 
			However, a problem with the charge to mass ratio still survives in this quantum field formulation.
			
		\end{abstract}

	\end{frontmatter}

	\section{Introduction}

	The original Kaluza-Klein idea \cite{Kaluza}\cite{Klein1}\cite{Klein2} consists in a $5D$ space-time formulation having the aim to include in a geometrical picture also the electromagnetic interaction. 
	
	\noindent The surprising formal success in providing a metric representation of the vector potential suggested, in the Seventies, to attempt for a geometrical unification \cite{ModernKKtheories}, able to assess all the fundamental interactions into a multi-dimensional space-time, with particular attention to the Electroweak Model. 
	The suggestive idea at the ground of these approach consists of the possibility to reproduce the Lie algebra, characterizing the elementary particle symmetries by the isometries of the extra-dimensional space. The non-trivial result obtained by the extra-dimensional Kaluza-Klein theories relies on the emergence from the multi-dimensional Einstein-Hilbert Lagrangian of the correct Yang-Mills action for the vector bosons which are the interaction carriers. 
	
	However, many non-trivial problems affected this fascinating attempt for a geometrization of Nature. One of the main questions came out from the difficulty to provide a geometrical version of the chirality singled out by the electroweak interaction \cite{Wetterich}, as well as the impossibility to represent the Standard Model of elementary particles in a Kaluza-Klein scenario \cite{Witten}.
	For alternative non-Riemannian approaches to solve the chirality problem of the Electroweak model see \cite{Cianfrani-Montani1}\cite{Cianfrani-Montani2}.

	\noindent Finally, we observe that a full geometrical picture of Nature would involve the geometrical formulation of the fermionic field, a really non-trivial perspective if supersymmetry is not considered \cite{Ferrara}. 
	
	\medskip
	
	Even the $5D$ Kaluza-Klein theory presents some important difficulties, see for a review \cite{Cianfrani-Marrocco-Montani}, which leaves open the question concerning the viability of this approach as a geometrization of the electromagnetic interaction. 
	
	\noindent First of all, the $5D$ metric tensor contains an additional degree of freedom besides the $4D$ metric and the vector potential, namely the fifth diagonal component. 
	Under the necessary restriction of the coordinate transformation in order to deal with the $U(1)$ symmetry, this quantity behaves as an additional scalar field, which presence non-trivially affects basic features of the electromagnetism, for instance, the charge conservation itself \cite{ModernKKtheories}\cite{Lacquaniti-Montani}\cite{Lacquaniti-Montani-Vietri}. 
	But, even fixing this scalar field to unity in the Lagrangian for the model (with the right sign of a space-like component), nonetheless, the ratio between the charge and the mass of an elementary particle is constrained to remain too small in order to reproduce the Standard Model spectrum of masses (for a proposal to solve the charge to mass ratio problem see \cite{Lacquaniti-Montani}).
	
	\noindent Finally, studying the morphology of a five-dimensional D'A\-lam\-ber\-tian\- operator, it is immediate to recognize the emergence of huge massive modes of a boson field, as result of the compactified scale of the fifth dimension \cite{Chodos-Detweiler}.
	
	\medskip 
	
	In the present analysis, we approach the formulation of the $5D$ Kaluza-Klein theory within the semi-classical and quantum framework of the so-called Polymer Quantum Mechanics \cite{Corichi1}\cite{Corichi2}.
	This revised formulation of quantum physics has the aim to introduce a discrete nature in the generalized coordinate (a real coordinate of a generic degree of freedom), as an effect of the emergence of cut-off physics. 
	
	Indeed, the fifth compactified dimension, being in the standard approach about two order greater than the Planck size, it is in the natural condition to be approached via the continuum limit of Polymer Quantum Mechanics as referred to a point particle living in this dimension.
	Furthermore, also the corresponding diagonal metric component (namely the additional Universe scale factor) in such a dynamical regime is to 
	be interested - as expected - by cut-off physics effects. 
	
	The present analysis follows the scenario proposed in \cite{Chodos-Detweiler} but revised in view of the polymer formulation. 
	
	\noindent We first show that a five-dimensional Kasner solution \cite{Kasner} \cite{Landau}\cite{Montani} (characterizing the Bianchi I Universe) admits a configuration in which three spatial directions isotropically expand, while the fourth remains static. 
	This result is of impact in the implementation of Kaluza-Klein theory, since it removes some of the non-trivial inconvenient features of a collapsing dimension, closely to a Planckian size. 
	For a previous attempt to deal with a static compactified dimension, on the base of a physical phenomenon, see \cite{Salam}. 
	
	\noindent Then, we analyse the geodesic motion on a generic $5D$ space-time, having a fifth steady dimension and we outline a natural solution to the charge to mass ratio problem. 
	This result comes out from the details of the semi-classical polymer formulation, adopted for the Hamiltonian dynamics of the free-falling particle. 
	In particular, the modified expression taken by the fifth momentum of the particle leads to a modified constitutive relation, that is - when passing from the momenta to the velocities - the previous constraint on the charge to mass ratio allows considering values which are natural in the Standard Model particles. 
	
	\noindent Finally, we study a five-dimensional Klein-Gordon equation and we clarify that, addressing the fifth coordinate via the quantum polymer prescription, the spectrum of emerging masses can fit some values of the Standard Model one and no tachyonic mode emerges, differently from the case discussed in \cite{Chodos-Detweiler}. 
	
	However, it should be noticed that, in this quantum field approach, a problem with the definition of a correct $q/m$ ratio for a Standard Model particle still survives.
	
	\medskip 
	
	The present study suggests that, when cut-off physics is included in the Kaluza-Klein formulation, some of the puzzling features of this approach are restated into a form that can give new physical insight for their understanding and overcoming. 
	
	\medskip 
	
	The manuscript presentation is structured as follows:
	in Section \ref{Kaluza-Klein theory} we review the main features of ordinary Kaluza-Klein theory, from the metric tensor construction and the resulting field equations to the geodesic motion of a point-like particle, which analysis, in particular, leads to the ordinary quantisation law for the electric charge, an estimate for the size $L$ of the fifth dimension and the aforementioned shortcoming
	of the charge to mass ratio of a particle. 
	
	\noindent In Section \ref{Polymer quantum mechanics} we review polymer quantum mechanics, summarizing the construction of the relative kinematics Hilbert space, via the introduction of a Weyl-Heisenberg algebra and under the assumption of the existence of a discrete spatial coordinate, and the implementation of the proper dynamics both on a quantum and semi-classical level, with particular regard to the p-polarization.
	
	\noindent In Section \ref{Polymer Kasner solution} we analyse the polymer-modified Kasner solution obtained from the introduction of the polymer framework on a semi-classical level in a $5D$ Bianchi I model, focusing on the behaviour of the fifth dimension.
	
	\noindent Finally, in Section \ref{Kaluza-Klein theory in polymer quantum mechanics framework}, based on the result of the previous section, first, we analyse, in a semi-classical formulation of Polymer Quantum Mechanics, the geodesic motion of a point-like particle and all its features, comparing all the results with the ones from the ordinary theory, and then we carry out the study of the polymer quantum dynamics of a complex Klein-Gordon field, along the lines of \cite{Chodos-Detweiler}, discussing with particular attention the resulting electric charge distribution and mass spectrum. 
	
	\noindent In Section \ref{Conlusion} brief concluding remarks follow.
	
	\section{Kaluza-Klein theory} 
	\label{Kaluza-Klein theory}
	
	Kaluza-Klein theory is a $5D$ extension of Einstein's General Relativity which aim is to provide a unified description of gravitational and electromagnetic interaction in a purely geometric fashion. 
	
	\noindent In the original theory \cite{Kaluza}\cite{Klein1}\cite{Klein2} the space-time is described by a $5D$ smooth manifold $V^5$, which is assumed to be the direct product $V^4\otimes S^1$ between a generic $4D$ manifold and a circus of length $L$, that is a compact space.
	
	\noindent A crucial assumption relies on the fact that all the observable physical quantities do not depend on the fifth coordinate $x^5$.
	This hypothesis can be further motivated by noticing that, due to the compactness of the fifth dimension, all the observable physical quantities are periodic in $x^5$; hence the independence on the fifth coordinate can be regarded as zero-order cut-off of a Fourier expansion of these quantities themselves, dubbed as cylinder condition.
	
	\noindent Once restricted the $5D$ general relativity principle to the following coordinate transformations (and their inverse):
	\begin{equation}   \label{KK_group_trans_1}
		\begin{cases}
			
			x^{\mu'}=\Psi(x^{\mu}) \\
			x^{5'}=x^5 + k\Lambda(x^{\mu}) \\
			
		\end{cases}
	\end{equation} 
	the $5D$ metric tensor of the expanded theory can be written as follows:
	\begin{equation}\label{KK_metric_tensor}
		\tilde{g}_{ab}=
		\left(
		\begin{array}{c|c}
			g_{\mu \nu} + k^2 \phi^2 A_\mu A_\nu & k \phi^2 A_\mu \\
			\hline
			k \phi^2 A_\nu & \phi^2
		\end{array}
		\right) ,
	\end{equation}
	where $g_{\mu \nu}$ is the $4D$ metric tensor of the ordinary theory, $A_{\mu}$ is the electromagnetic four-potential, $\phi$ is a scalar field and $k$ is a constant to be properly determined. 
	
	\subsection{Kaluza-Klein field equations}
	The field equations of the theory can be obtained from a $5D$ Einstein-Hilbert action:
	\begin{equation}
		^{(5)}S:=\tilde{S}= - \frac{1}{16\pi\tilde{G}}\int_{V^{4}\otimes S^1}  dx^{0}dx^{1}dx^{2}dx^{3}dx^{5}   \sqrt{-\tilde{g}} \tilde{R} ,
	\end{equation}
	where $\tilde{G}$, $\tilde{g}$ and $\tilde{R}$ are respectively the $5D$ gravitational constant, the metric tensor $\tilde{g}_{ab}$ determinant and the $5D$ scalar curvature.
	
	By performing a 4+1 dimensional reduction the ordinary $4D$ Einstein-Maxwell action is surprisingly obtained:
	\begin{equation}
		\begin{split}
			\tilde{S}=& - \frac{c^3}{16\pi{G}}\int_{V^{4}}  d^4x \sqrt{{-g}}\phi \biggl( R + \frac{1}{4}\phi^2 k^2 F_{\mu \nu}F^{\mu \nu}\\
			&+\frac{2}{\phi}\nabla_{\mu} \partial^{\mu} \phi \biggr).
		\end{split}
	\end{equation}
	By setting $\phi=1$ in the action - as in the original work of Kaluza \cite{Kaluza} and Klein \cite{Klein1}\cite{Klein2} - and by using the variational principle ordinary, Einstein-Maxwell field equations can be correctly recovered, once $k$ is setted equal to $2\sqrt{G}/c^2$.

	\subsection{Geodesic motion}
	
	A free point-like particle in this theory will move along a $5D$ geodesic, hence the respective action, with signature (-,+,+,+,+), will be:
	\begin{equation}
		\tilde{S}=-mc\int d\tilde{s}=-mc \int \sqrt{-\tilde{g}_{a b}\frac{d x^a}{d \tilde{s}}
			\frac{d x^b}{d \tilde{s}}} d\tilde{s},
	\end{equation}
	where $d\tilde{s}$ is the $5D$ line element, to be distinguished from the $4D$ line element $ds$.
	
	\noindent Once setted - here and in the further developments - $\phi=1$ in the metric \eqref{KK_metric_tensor}, from the variational principle $5D$ geodesic equation is immediately restored:
	\begin{equation} \label{geodesic}
		\tilde{u}^a \tilde{\nabla}_{a} \tilde{u}^b=0.
	\end{equation}
	
	It is essential to point out that $5D$ velocity $\tilde{u}^a$ is different from $4D$ velocity $u^a$; indeed they are related as follows:
	\begin{equation} \label{KK_five-four_velocity_rel}
		\tilde{u}^a=\frac{1}{\sqrt{1-u_5^2}} u^a.
	\end{equation}
	
	\noindent From relations \eqref{geodesic} and \eqref{KK_five-four_velocity_rel} it can be easily shown that $u_5$ is a constant of motion. 
	
	In order to achieve $4D$ equation of motion, the geodesic equation \eqref{geodesic} has to be evaluated for the usual space-time variables only, which we indicate with Greek letters. 
	
	\noindent By making use of relation \eqref{KK_five-four_velocity_rel}, the following result is attained:
	\begin{equation}
		u^\nu\nabla_{\nu}u^{\mu}=\frac{2\sqrt{G}}{c^2}u_5u^{\nu}g^{\mu \lambda}F_{ \nu \lambda},
	\end{equation}
	where $F_{ \nu \lambda}$ is the antisymmetric electromagnetic tensor. 
	
	\noindent By comparison with the ordinary classical equation:
	\begin{equation} \label{KK_ordinary_4d_geodesic}
		u^\nu\nabla_{\nu}u^{\mu}=\frac{q}{mc^2}u^{\nu}g^{\mu \lambda}F_{\nu\lambda},
	\end{equation}
	the following fundamental identification is achieved:
	\begin{equation} \label{KK_u5_rel_q}
		u_{5}=\frac{q}{2m\sqrt{G}}.
	\end{equation}
	\noindent Being $p_5=mcu_5$ it can then be written:
	\begin{equation} \label{KK_p5_rel_q}
		p_5=\frac{qc}{2 \sqrt{G}},
	\end{equation}
	which establishes a fundamental relation between the particle fifth component of momentum and its electric charge. 
	
	The compactness of the fifth dimension implies a quantisation of momentum along the fifth direction:
	\begin{equation} \label{KK_p_quantised}
		p_5=\frac{2\pi \hbar}{L}n \qquad n \in \mathbb{Z},
	\end{equation}
	where we remind that $L$ is the length of the circus describing the fifth dimension.
	\noindent By a direct comparison between relations \eqref{KK_p5_rel_q} and \eqref{KK_p_quantised} a natural quantisation law for the electric charge and an estimate of the size $L$ of the fifth dimension are obtained:
	\begin{equation} \label{KK_ordinary_L}
		{L}=4\pi \frac{\hbar\sqrt{G}}{e c}\approx 2.37\cdot10^{-31}\>cm \qquad q=ne,
	\end{equation}
	where $e$ is the electron charge.
	
	\noindent Coherently the size of the fifth dimension is in agreement with its non-observability and with its impossibility to be currently detected. 
	
	\medskip 
	
	Nevertheless, despite these remarkable results, the relation \eqref{KK_five-four_velocity_rel} sets the constraint $\abs{u_5}<1$; 
	by virtue of relation \eqref{KK_u5_rel_q}, this implies the following condition on the charge/mass ratio of a particle:
	\begin{equation} \label{KK_q_over_m}
		\frac{\abs{q}}{m}<2\sqrt{G}\approx 5.16\cdot10^{-4} \> e.s.u./g,
	\end{equation}
	which, unfortunately, has not phenomenological confirmation, neither for elementary particle nor for macroscopic object, hence representing one of the puzzling shortcomings of the theory.
	
	\section{Polymer quantum mechanics}
	\label{Polymer quantum mechanics}
	
	Polymer quantum mechanics is a non-standard representation of the non-relativistic quantum theory, unitarily inequivalent to the Schrödinger one \cite{Corichi1}\cite{Corichi2}.
	Its developments are due mainly to the exploration of background-independent theories, such as quantum gravity, of which mimics several structures \cite{Ashtekar}. 
	
	\smallskip
	
	Given a discrete orthonormal basis $\ket{\mu_i}$ for a space $\mathcal{H'}$, such that $\braket{\mu_i}{\mu_j}=\delta_{ij}$, where $\mu_i \in \mathbb{R}$ and $i=1,2...,n$, the kinematic polymer Hilbert space $\mathcal{H}_{poly}$ is obtained as a Cauchy completion of $\mathcal{H'}$.
	
	\noindent On this space two abstract operators can be defined:
	\begin{align}
		& \hat{\epsilon}\ket{\mu}:=\mu \ket{\mu} \\
		& \hat{s}(\lambda)\ket{\mu}:=\ket{\mu +\lambda}.
	\end{align}
	\noindent The operator $\hat{\epsilon}$ is a symmetric operator and $\hat{s}(\lambda)$ defines a one-parameter family of unitary operators.
	In spite of this $\hat{s}(\lambda)$ is discontinuous with respect to $\lambda$;
	this means that no self-adjoint operator exists that can generate $\hat{s}(\lambda)$ by exponentiation. 
	Taking now in exam a physical system with configuration space spanned by the coordinate $q$, which is assumed to have a discrete character, and its conjugate momentum $p$, the previous abstract representation can be projected and studied with respect to p-polarization.
	In this polarization the basis states will be:
	\begin{equation} \label{basic_states_p}
		\psi_\mu(p)=\braket{p}{\mu}=e^{i\mu p/\hbar}.
	\end{equation}
	Following the algebraic construction method, a Weyl-Hei\-sen\-berg\- algebra is introduced on $\mathcal{H}_{poly}$ and the action of its generators on the basis states is defined as follows:
	\begin{align}
		&\hat{\mathcal{U}}(\nu)\psi_\mu(p)=\psi_\mu(p+\nu)=e^{i\mu(p+\nu)/\hbar}=e^{i\mu\nu /\hbar}e^{i\mu p/\hbar} \\
		&\hat{\mathcal{V}}(\lambda)\psi_\mu(p)=e^{i\lambda p/\hbar}e^{i\mu p/\hbar}=e^{i(\lambda + \mu)p/\hbar}=\psi_{\mu+\lambda}(p).
	\end{align}
	From this it can be inferred that the shifting operator $\hat{s}(\lambda)$ can be identified with the operator $\hat{\mathcal{V}}(\lambda)$, which is then discontinuous in $\lambda$; this means that the spatial translations generator, that is the momentum operator $\hat{p}$, does not exist. 
	On the other hand, the operator $\hat{\mathcal{U}}(\nu)$ is continuous, so that the translations generator in the momentum space, i.e. the position operator $\hat{q}$, exists and it can be identified with the abstract operator $\hat{\epsilon}$.
	
	\noindent Indeed:
	\begin{equation} \label{PQ_q_op}
		\hat{q}\psi_{\mu}(p)=-i\hbar \partial_p\psi_{\mu}(p)=\mu\psi_{\mu}(p).
	\end{equation}
	It can be proved \cite{Corichi2} that the kinematic polymer Hilbert space in this polarization is explicitly given by $\mathcal{H}_{poly,p}=L^2\left(\mathbb{R}_{B},d\mu_{H}\right)$, where $\mathbb{R}_{B}$  is the so-called Bohr compactification of the real line and $d\mu_{H}$ is the Haar measure. 
	
	\medskip 
	
	A similar picture is obtained in the q-polarization: the momentum operator cannot still be defined, while it is possible to show that the fundamental wave functions are Kronecker deltas and that the kinematic polymer Hilbert space is explicitly given by $\mathcal{H}_{poly,x}=L^2\left(\mathbb{R}_{d},d\mu_{c}\right)$, where $\mathbb{R}_{d}$  is the real line equipped with a discrete topology and $d\mu_{c}$ is the counting measure. 
	
	\medskip 
	
	In order to build the dynamics a Hamiltonian operator $\hat{H}$ has to be defined on $\mathcal{H}_{poly}$, but since $\hat{p}$ do not exist, a direct implementation is not possible.
	To overcome this problem the momentum operator can be approximated by defining on the configuration space of the system a regular graph $\gamma_{\mu}=\{q \in \mathbb{R} \> | \> q=n\mu, n \in \mathbb{Z}\}$, where $\mu$ is the fundamental scale introduced by the polymer representation. 
	The basis kets $\ket{\mu}$ can now be indicated as $\ket{\mu_n}$, where $\mu_n=n\mu$ are the points belonging to the graph $\gamma_{\mu_0}$.
	Consequently the generic states will be:
	\begin{equation}
		\ket{\psi}_{\gamma_{\mu}}=\sum_n a_n \ket{\mu_n},
	\end{equation}
	and they will belong to the new Hilbert space $\mathcal{H}_{\gamma_{\mu}} \subset \mathcal{H}_{poly}$, posed that they satisfy the condition $\sum_n \abs{a_n}^2 < \infty$.
	Since the dynamics has to be closed in $\mathcal{H}_{\gamma_{\mu}}$, the shift parameter $\lambda$ has to be fixed equal to $\mu$, hence the action of $\hat{\mathcal{V}}(\lambda)$ will be:
	\begin{equation}
		\hat{\mathcal{V}}(\lambda)\ket{\mu_n}=\hat{\mathcal{V}}(\mu)\ket{\mu_n}=\ket{\mu_{n+1}}.
	\end{equation}
	
	\noindent On a general ground, the variable $p$ it can be written as:
	\begin{equation} \label{PQ_p_var_approx}
		p \approx \frac{\hbar}{\mu} \sin(\frac{\mu}{\hbar}p)=\frac{\hbar}{2i\mu}\left(e^{i\frac{\mu}{\hbar}p}-e^{-i\frac{\mu}{\hbar}p}\right),
	\end{equation}
	when the condition $p<<\hbar/\mu$ holds.
	
	\noindent Based on this approximation and visualizing the action of  $\hat{\mathcal{V}}(\mu)$ in the p-polarization, it is clear that the operator $\hat{p}$ and its action can be approximated as:
	\begin{equation} \label{PQ_p_op_approx}
		\hat{p}_{\mu}\ket{\mu_n}\approx \frac{\hbar}{2i\mu} \left[\hat{\mathcal{V}}(\mu)-\hat{\mathcal{V}}(-\mu)\right]\ket{\mu_n},
	\end{equation}
	where $\mu$ acts as a regulator.
	
	\noindent To approximate the operator $\hat{p}^2$ two paths are possible:
	\begin{equation} \label{PQ_p^2_op_approx_1}
		\hat{p}^2_{\mu}\approx \frac{\hbar^2}{4\mu^2} \left[ 2-\hat{\mathcal{V}}(2\mu)-\hat{\mathcal{V}}(-2\mu)\right],
	\end{equation} 
	based on the approximation 
	\begin{equation} \label{PQ_p^2_var_approx_1}
		p^2 \approx \frac{\hbar^2}{\mu^2} \sin[2](\frac{\mu}{\hbar}p)
	\end{equation}
	and hence defined by iterating the action of $\hat{p}$ according \eqref{PQ_p_op_approx}, or
	\begin{equation} \label{PQ_p^2_op_approx_2}
		\hat{p}^2_{\mu}\approx \frac{\hbar^2}{{\mu}^2} \left[2-\hat{\mathcal{V}}(\mu)-\hat{\mathcal{V}}(-\mu)\right],
	\end{equation}
	exploiting the approximation 
	\begin{equation}\label{PQ_p^2_var_approx_2}
		p^2 \approx \frac{2\hbar^2}{{\mu}^2}\left(1- \cos(\frac{\mu}{\hbar}p)\right),
	\end{equation}
	valid as long as $p<<\hbar/\mu$.

	\noindent Hence, the well-defined, symmetric Hamiltonian operator will be:
	\begin{equation}
		\hat{H}_{\mu}:=\frac{\hat{p}^2_{\mu}}{2m}+\hat{V}(\hat{q}),
	\end{equation}
	where $\hat{V}(\hat{q})$ is the potential operator. 
	
	Therefore, quantising a system according to the polymer representation, in the p-polarization, implies the use of the approximation \eqref{PQ_p^2_op_approx_1} or \eqref{PQ_p^2_op_approx_2} for the momentum operator, while the position operator will be the natural differential operator, which action is expressed in \eqref{PQ_p_op_approx}.
	
	In a semi-classical approach, this procedure corresponds to the proper introduction of the approximations \eqref{PQ_p_var_approx}, \eqref{PQ_p^2_var_approx_1} and \eqref{PQ_p^2_var_approx_2} on the variable $p$ in the dynamics of the system of interest. 
	Hence, on this level, the whole procedure can be thought of as a prescription to provide physical insight into the behaviour of the quantum expectations values, according to the so-called Ehrenfest theorem.

	\section{Polymer Kasner solution}
	\label{Polymer Kasner solution}
	
	We will now apply polymer formalism in a semi-classical framework to the study of a $5D$ Bianchi I model, which solution in the vacuum is the well-known Kasner metric \cite{Kasner}\cite{Landau}, focusing on the kinematics and dynamics of the fifth dimension. 
	
	\medskip
	
	\noindent In order to obtain the polymer Kasner cosmological solution, we need a minisuperspace Hamiltonian formulation, extended to the $5D$ case.
	
	\noindent The $5D$ Bianchi I line element, written in the ADM formalism \cite{ADM}, is a straightforward generalisation of the $4D$ one\footnote{As it should be clear from the context, the metric coefficient $c(t)$ is not to be confused with the velocity of light $c$.}:
	\begin{equation}
		\begin{split}
			ds^2=&-N^2(t)c^2dt^2+^{(4)}h_{ij}dx^idx^j=\\
			&-N^2(t)c^2dt^2+a^2(t)(dx^1)^2+b^2(t)(dx^2)^2\\
			& +c^2(t)(dx^3)^2+d^2(t)(dx^5)^2, \qquad (\text{i,j=1,2,3,5})
		\end{split}
	\end{equation}
	where $N(t)$ is the lapse function of the ADM formalism, $^{(4)}h_{ij}$ is the metric tensor of the $4D$ manifold, which coordinate are all space-like.
	
	\noindent Having the general structure of this metric as starting point, we can build the Hamiltonian of the system:
	\begin{equation} \label{PBM_hamiltonian_bianchi_non_diag}
		\begin{split}
			H_{Bianchi \> I}:=H_{B}=Ne^{-\sum_a q^a/2}\biggl\{\sum_a p_a^2-\frac{1}{3}\biggl[\sum_b p_b\biggr]^2\biggr\},
		\end{split}
	\end{equation}
	which, by varying with respect to $N(t)$ turns out to be a constraint for the dynamics, namely $H_B=0$.
	
	\noindent The couple $(q^a,p_a)$  in \eqref{PBM_hamiltonian_bianchi_non_diag} are the conjugate variables spanning an highly symmetric phase space, the so-called \textit{minisuperspace}, and the relation between the metric coefficients and the q-variables is the usual one of the literature \cite{Montani}, extended to the $5D$ case:
	\begin{equation} \label{PBM_metric_coeff_to_q}
		\begin{split}
			&a(t)=e^{q^1(t)/2} \quad b(t)=e^{q^2(t)/2} \\
			&c(t)=e^{q^3(t)/2} \quad d(t)=e^{q^5(t)/2}.
		\end{split}
	\end{equation}
	
	As is well-known, it is more convenient to express the obtained Hamiltonian in its diagonal form:
	\begin{equation} \label{PBM_hamilt_bianchi_diagonal}
		H'_{B}=Ne^{-\alpha}\left[-\frac{1}{3}p^2_\alpha+p^2_{+}+p^2_{-}+p^2_{\gamma}\right],
	\end{equation}
	which is the canonical form of the quadratic form associated to $H_{B}$. 
	
	\noindent The p-variables in the Hamiltonian \eqref{PBM_hamilt_bianchi_diagonal} are the conjugate momenta of a set of variables $\alpha,\beta_{+},\beta_{-},\gamma$, which represent the generalisation of the Misner variables \cite{Misner}. The relation between these variables and the previous q-variables is defined through the following linear transformation:
	\begin{equation} \label{q_to_misner_var}
		\begin{cases}
			q^1=\frac{1}{2}\alpha-\frac{1}{2\sqrt{3}}\beta_{+}-\frac{1}{\sqrt{6}}\beta_{-}-\frac{1}{\sqrt{2}}\gamma\\ q^2=\frac{1}{2}\alpha-\frac{1}{2\sqrt{3}}\beta_{+}-\frac{1}{\sqrt{6}}\beta_{-}+\frac{1}{\sqrt{2}}\gamma \\
			q^3=\frac{1}{2}\alpha-\frac{1}{2\sqrt{3}}\beta_{+}+\sqrt{\frac{2}{3}}\beta_{-} \\
			q^5=\frac{1}{2}\alpha+\frac{\sqrt{3}}{2}\beta_{+}.
		\end{cases}
	\end{equation}

	By using the Hamilton-Jacobi method and the Hamilton equation for the variable $\alpha$ - which represents the universe volume - in the synchronous reference frame, the standard classical Kasner solution for the $5D$ case can be recovered:
	\begin{equation} \label{PBM_kasner_metric}
		\begin{split}
			ds^2&=-c^2dt^2+(t/t_0)^{2k_1}(dx^1)^2+(t/t_0)^{2k_2}(dx^2)^2\\
			&+(t/t_0)^{2k_3}(dx^3)^2+(t/t_0)^{2k_5}(dx^5)^2.
		\end{split}
	\end{equation}
	
	\noindent The k parameters are the so-called Kasner exponents and they satisfy the following conditions:
	\begin{equation} \label{PBM_kasner_buondaries_standard}
		\begin{cases}
			k_1+k_2+k_3+k_5=1 \\
			k_1^2+k_2^2+k_3^2+k_5^2=1.
		\end{cases}
	\end{equation}
	\noindent In particular, if we assume isotropy in the three usual spatial dimensions, as observations suggest, that is if we set $k_1=k_2=k_3$, solution of the previous system becomes:
	\begin{equation}
		k_1=k_2=k_3=\frac{1}{2} \quad k_5=-\frac{1}{2}.
	\end{equation}
	\noindent This means that while the three usual spatial dimensions expand, the fifth one collapses indefinitely. 
	
	\medskip 
	
	We want now to introduce polymer formalism in Bianchi I dynamics.
	In order to do this we choose to operate the substitutions \eqref{PQ_p_var_approx} and \eqref{PQ_p^2_var_approx_1} on the conjugate momentum $p_\gamma$ of the Misner variable $\gamma$, connected with the metric coefficient of the fifth dimension.
	The new \textit{polymerized} Hamiltonian will be:
	\begin{equation} \label{PBM_hamlit_poly_Bianchi}
		H_{B}^{poly}=e^{-\alpha}\biggl[-\frac{1}{3}p^2_\alpha+p^2_{+}+p^2_-+\frac{\hbar^2}{\mu^2}\sin[2](\frac{\mu}{\hbar}p_\gamma)\biggr].
	\end{equation}
	
	\noindent By the exactly above procedure we find that the solution is still a Kasner solution, where the metric coefficients have a coordinate time power trend as in \eqref{PBM_kasner_metric}, but their exponents, that is the Kasner indices, due to the quantum polymer modifications, satisfy different constraints:
	
	\begin{equation}
		\begin{cases}
			k_1+k_2+k_3+k_5=1 \\
			k_1^2+k_2^2+k_3^2+k_5^2=1-\frac{3}{4}\frac{\hbar^2}{\mu^2}\frac{\sin[4](\frac{\mu}{\hbar}p_{\gamma})}{\sqrt{p^2_++p^2_-+\frac{\hbar^2}{\mu^2}\sin[2](\frac{\mu}{\hbar}p_\gamma)}}.
		\end{cases}
	\end{equation}
	
	\noindent The second term of the right-hand side of the second condition is non-negative, so that we can restate the system as follows:
	\begin{equation}
		\begin{cases} \label{PBM_kasner_ind_poly_conditions}
			k_1+k_2+k_3+k_5=1 \\
			k_1^2+k_2^2+k_3^2+k_5^2\leq 1.
		\end{cases}
	\end{equation}
	
	\noindent Assuming isotropy in the three usual spatial dimensions and introducing an order between exponents, in particular setting $k_5<k$, the system \eqref{PBM_kasner_ind_poly_conditions} has the following solution:
	\begin{equation} \label{PBM_constraint_polymer_Kasner}
		\begin{cases}
			1/4 < k\leq1/2 \\
			-1/2 \leq k_5 < 1/4.
		\end{cases}
	\end{equation}
	
	We can notice that the Kasner exponents relative to the three usual spatial dimensions are bound to be positive and they can take value in a precise interval, while the exponent relative to the fifth dimension has value in an interval in which it can take both positive and negative values, which is a new feature of the Kasner solution entirely due to the polymer physics.
	
	\noindent The latter statement grows in importance when we draw our attention to the permitted values of $k_5$.
	
	\medskip
	
	Indeed, even if the power trend with respect to the coordinate time variable of fifth dimension metric coefficient - which is ultimately responsible for the singularity at an infinite time - is not removed in this scenario, the introduction of the polymer formalism leads to a modification on the constraint of the fifth Kasner exponent $k_5$, which interval of definition allows the value $k_5=0$.

	\noindent For this particular choice of $k_5$, the other Kasner indices will be equal to $k=1/3$, correctly reproducing the observed isotropic expansion in the three usual spatial dimensions, while the metric coefficient relative to the fifth dimension will be of course equal to one. 
	
	\noindent This means that the fifth dimension has no dynamics since the time-dependence of the relative metric coefficient disappears. 
	We, therefore, have obtained a static solution, which somehow solves the singularity problem by remarkably removing the indefinite collapse of the dimension itself.
	
	\noindent The developments of the next sections will be based on this important result.
	
	\section{Kaluza-Klein theory in polymer quantum mechanics framework}
	\label{Kaluza-Klein theory in polymer quantum mechanics framework}
	
	In this section we face the analysis of the Kaluza-Klein paradigm in the framework of Polymer Quantum Mechanics, both studying the behaviour of the geodesic motion and the quantum dynamics of a Klein-Gordon field. 
	From the point of view of the geodesic discussion, the polymer framework is addressed on a semi-classical level, in the spirit of the Ehrenfest theorem. 
	Instead, the study of the quantum scalar field is performed in a full quantum picture, as restricted to the fifth coordinate. 
	
	\subsection{Geodesic motion}
	
	In this section polymer formalism will be now applied to the Hamiltonian formulation of the $5D$ geodesic motion, at a semi-classical level, that is introducing quantum modifications to the classical dynamics. 
	
	\medskip
	
	The $5D$ Hamiltonian of a free particle in a general Kaluza-Klein background, with $\tilde{g}_{55}=1$ - see previous section - reads as:
	\begin{equation}
		\tilde{H}=\frac{1}{2mc}\left[\tilde{p}_{\mu} \tilde{p}_{\nu} {g}^{\mu\nu}-\frac{4\sqrt{G}}{c^2}\tilde{p}_\mu \tilde{p}_5A^{\mu}+\tilde{p}^2_5\left(1+\frac{4G}{c^4}A_{\mu}A^{\mu} \right)\right].
	\end{equation}
	where, as before, the quantities with a tilde are the $5D$ ones, that is those defined with respect to the $5D$ line element.
	
	We want now introduce polymer formalism only with respect to the canonical couple $(x^5,\tilde{p}_5)$, that is we assume that the coordinate $x^5$ has an essential discrete nature and we redefine the respective conjugate momentum $\tilde{p}_5$ by introducing a regular graph structure on $S^1_{x^5}$, i.e. on the fifth dimension, which is now equipped with a discrete topology, according to the discussion in Section \ref{Polymer quantum mechanics}.
	Following the polymer prescription \eqref{PQ_p_var_approx}, we can made the substitution:
	\begin{equation}
		\tilde{p}_5 \rightarrow \frac{\hbar}{\mu}\sin(\frac{\mu}{\hbar}\tilde{p}_5).
	\end{equation}
	
	\noindent The new Hamiltonian will be rewritten as:
	\begin{equation}
		\begin{split}
			\tilde{H}_{poly}= & \frac{1}{2mc}\bigg[\tilde{p}_{\mu} \tilde{p}_{\nu} {g}^{\mu\nu}-\frac{4\sqrt{G}}{c^2}\frac{\hbar}{\mu}\tilde{p}_\mu \sin(\frac{\mu}{\hbar}\tilde{p}_5)A^{\mu}\\
			&+\frac{\hbar^2}{\mu^2}\sin[2](\frac{\mu}{\hbar}\tilde{p}_5)\left(1+\frac{4G}{c^4}A_{\mu}A^{\mu} \right)\bigg],
		\end{split}
	\end{equation}
	
	from which the equations of motion (Hamilton equations) can be obtained:
	\begin{equation} \label{PG_polymer_equation_of_motion_4_index}
		\begin{cases}
			\tilde{u}^\mu=\frac{1}{mc}\left[\tilde{p}_{\nu}{g}^{\mu \nu}-\frac{2\sqrt{G}}{c^2}\frac{\hbar}{\mu}\sin(\frac{\mu}{\hbar}\tilde{p}_5)A^{\mu}\right] \\
			\begin{split}
				\dot{\tilde{p}}_{\mu}=&-\frac{1}{2mc}\bigg[\tilde{p}_{\rho}\tilde{p}_{\sigma}\partial_{\mu}{g}^{\rho \sigma}-\frac{4\sqrt{G}}{c^2}\tilde{p}_{\rho}\frac{\hbar}{\mu}\sin(\frac{\mu}{\hbar}\tilde{p}_5)\partial_{\mu}A^{\rho}\\
				&+\frac{4G}{c^4}\frac{\hbar^2}{\mu^2}\sin[2](\frac{\mu}{\hbar}\tilde{p}_5)\partial_{\mu}\left(A_\nu A^\nu\right)\bigg], \\
			\end{split}
		\end{cases}
	\end{equation}
	
	\begin{equation} \label{PG_polymer_equation_of_motion_5}
		\begin{cases}
			\begin{split}
				\tilde{u}^5=&\frac{1}{mc}\bigg[-\frac{2\sqrt{G}}{c^2}\tilde{p}_\mu A^{\mu}\cos(\frac{\mu}{\hbar}\tilde{p}_5)\\
				&+\frac{\hbar}{\mu}\sin(\frac{\mu}{\hbar}\tilde{p}_5)\cos(\frac{\mu}{\hbar}\tilde{p}_5)\left(1+\frac{4G}{c^4}A_\nu A^\nu\right)\bigg] 
			\end{split} \\	 
			\dot{\tilde{p}}_5=0,
		\end{cases}
	\end{equation}
	where the dot refers to the derivative with respect to the five-dimensional line element $d\tilde{s}$. 
	
	\noindent We outline that in the last equation we made use of the \textit{cylinder condition}, which implies that $\tilde{p}_5$ is a constant of motion, as expected.
	
	\noindent As we know from the standard case, in order to make a comparison with the $4D$ equations of motion, we need to know the relation between $4D$ and $5D$ quantities, in particular the relation between five-velocities and four-velocities.
	These are formally the same of equation \eqref{KK_five-four_velocity_rel} also in our framework.
	
	Nevertheless, while in the ordinary theory, with the scalar field of the Kaluza Klein metric set to be constant, $u_5$ is a constant of motion, in the polymer framework this is not the case. 
	
	\noindent Indeed, from the explicit expression of $\tilde{u}^5$ \eqref{PG_polymer_equation_of_motion_5} we can obtain the corresponding expression of $\tilde{u}_5$:
	
	\begin{equation}
		\tilde{u}_5=\frac{2\sqrt{G}}{c^2} \tilde{u}^\mu A_\mu \cos(\frac{\mu}{\hbar}\tilde{p}_5)+\frac{\hbar}{\mu mc}\sin(\frac{\mu}{\hbar}\tilde{p}_5)\cos(\frac{\mu}{\hbar}\tilde{p}_5),
	\end{equation}
	
	where we have made use of the first Hamilton equation for $\tilde{u}^{\mu}$.
	
	\noindent The latter can be solved in function of $u_5$ and two solutions are obtained:
	\begin{equation} \label{PG_u_5 exact expression}
		\begin{split}
			u_5=&\frac{1}{D(\mu,\tilde{p}_5)}\bigg\{16\sqrt{G}  \mu ^2 m^2   \left(1-\cos(\frac{\mu  }{\hbar}\tilde{p}_5)\right)A_\mu u^\mu\\
			&\pm \sqrt{2}\hbar\abs{\sin(2\frac{\mu }{\hbar}\tilde{p}_5)} \sqrt{D(\mu, \tilde{p}_5)+ K(\mu,\tilde{p}_5)A_\mu A^\mu u_\nu u^\nu}\bigg\},
		\end{split}
	\end{equation}
	where:
	\begin{gather}
		D(\mu, \tilde{p}_5):={\hbar^2+8\mu ^2 m^2 c^2-\hbar^2\cos (4\frac{\mu  }{\hbar}\tilde{p}_5)} \\
		K(\mu,\tilde{p}_5):=-128 \frac{G}{c^2} \mu ^2 m^2\sin[4](\frac{\mu}{2\hbar}\tilde{p}_5).
	\end{gather}
	
	\noindent These are rather complicated non-constant expressions, due to the dependence from the electromagnetic and the four-velocity fields, which consequently heavily modify the $\frac{ds}{d\tilde{s}}$ factor. 
	
	\noindent From the Hamilton equations \eqref{PG_polymer_equation_of_motion_4_index}, by exploiting the formal relation \eqref{KK_five-four_velocity_rel}, we can write down the part relative to the $4D$ indices of the new generalised $5D$ geodesic, as a function of the $4D$ quantities:
	
	\begin{equation} \label{PG_4d_polymer_geodesic_equation}
		u^{\nu}\nabla_{\nu}u^{\mu}=\frac{2\sqrt{G}\hbar}{mc^3\mu} \sqrt{1-u_5^2}\sin(\frac{\mu}{\hbar}\tilde{p}_5)u^{\nu}F^{\mu}_{\nu}-\frac{u_5u^{\mu}}{1-u^2_5}\frac{du_5}{ds},
	\end{equation}
	in which it is possible to distinguish the polymer-modified part of the standard $4D$ geodesic \eqref{KK_ordinary_4d_geodesic} and an extra term, entirely due to the variation of $u_5$. 
	It is worth pointing out that eq.\eqref{PG_4d_polymer_geodesic_equation} correctly reduces to the standard one in the $\mu \to 0$ limit.
	
	\noindent At this point, in order to compare the last expression with the equation of motion \eqref{KK_ordinary_4d_geodesic} we need to make a series expansion of $u_5$ by assuming $A_\mu << 1$.
	This can be perfectly legitimate in a cosmological setting since there is no 
	relevant coherent electromagnetic field which permeates the Universe.
	
	This leads to the following expression, to the first order in $A_\mu$:
	\begin{equation}
		\begin{split}
			u_5\approx &\pm\bigg\{\frac{\sqrt{2} \hbar\abs{\sin(2\frac{\mu}{\hbar}\tilde{p}_5) }}{D(\mu,\tilde{p}_5)^{1/2}}+\frac{16\sqrt{G}\mu^2m^2 \left(\cos(\frac{\mu }{\hbar}\tilde{p}_5) -1\right)}{D(\mu,\tilde{p}_5)} A_{\mu}u^{\mu}\\ 
			&+O\left(A^2\right)\bigg\}.
		\end{split}
	\end{equation}
	\noindent Hence, by inserting this expansion in the geodesic equation \eqref{PG_4d_polymer_geodesic_equation} and ignoring the extra term - which meaning it is beyond the purpose of this article - we obtain, keeping only the overall first order term in $A_\mu$:
	\begin{equation}
		u^{\nu}\nabla_{\nu}u^{\mu}\approx\frac{2\sqrt{G}\hbar}{\mu mc^3} \frac{2\sqrt{2}\mu mc \sin(\frac{\mu}{\hbar}\tilde{p}_5)}{\sqrt{\hbar^2+8\mu^2m^2c^2-\hbar^2 \cos(4\frac{\mu}{\hbar}\tilde{p}_5)}}u^{\nu}F^{\mu}_{\nu}
	\end{equation}
	from which we can read off the perturbative polymer relation between the electric charge $q$ and $\tilde{p}_5$ - which we remind it is a constant of motion:
	\begin{equation} \label{PG_polymer_q_expression}
		q\approx\frac{2\sqrt{G}}{c} \frac{2\sqrt{2}\hbar mc }{\sqrt{\hbar^2+8\mu^2m^2c^2-\hbar^2 \cos(4\frac{\mu}{\hbar}\tilde{p}_5)}}\sin(\frac{\mu}{\hbar}\tilde{p}_5).
	\end{equation}
	
	To correctly interpret this formula we can refer to the ordinary relation between $q$ and $\tilde{p}_5$, which can be obtained from the expression \eqref{KK_p5_rel_q}.
	In fact, since in the standard case the following simple statement holds:
	\begin{equation} \label{PG_ordinary_rel_between_p_5}
		\tilde{p}_5=\frac{mc}{\sqrt{m^2c^2-p^2_5}}p_5 \qquad 
		p_5=\frac{mc}{\sqrt{m^2c^2+\tilde{p}^2_5}}\tilde{p}_5,
	\end{equation}
	the relation \eqref{KK_p5_rel_q} can be rewritten as:
	\begin{equation} \label{PG_rel_q_p5_standard}
		q=\frac{2\sqrt{G}}{c}\frac{mc}{\sqrt{m^2c^2+\tilde{p}^2_5}}\tilde{p}_5.
	\end{equation}
	\noindent This suggests that the expression \eqref{PG_polymer_q_expression} is the polymer generalisation to the first perturbative order in the electromagnetic field of the latter relation.
	
	\noindent This interpretation is confirmed by the $\mu\to0$ limit of \eqref{PG_polymer_q_expression}, from which the standard expression \eqref{PG_rel_q_p5_standard} is recovered. 
	
	\noindent Therefore, based on the relations \eqref{PG_ordinary_rel_between_p_5}, we can impose:
	\begin{equation}
		\frac{\hbar}{\mu}\sin(\frac{\mu}{\hbar}p_5)\approx\frac{2\sqrt{2}\hbar mc }{\sqrt{\hbar^2+8\mu^2m^2c^2-\hbar^2 \cos(4\frac{\mu}{\hbar}\tilde{p}_5)}}\sin(\frac{\mu}{\hbar}\tilde{p}_5)
	\end{equation}
	and, in the end:
	\begin{equation} \label{PG_poly_charge_p5}
		q\approx\frac{2\hbar\sqrt{G}}{\mu c}\sin(\frac{\mu}{\hbar}p_5).
	\end{equation}
	\noindent This is a periodic and bounded function of $p_5$ with period $2\pi$, defined in the generic periodic interval $[\pi(2k-1)\hbar/\mu,\pi(2k+1)\hbar/\mu]$, with $k\in\mathbb{Z}$. 
	Since we want to preserve the Cauchy problem symmetry with respect to the initial values of $p_5$, that is the symmetry of left-handed and right-handed particles in the fifth direction, we are led to choose the interval $[-\pi\hbar/\mu,\pi\hbar/\mu]$, i.e. the one for k=0, as the natural periodic interval of definition.
	In the $\mu\to0$ limit, the obtained function correctly reproduces the ordinary relation inferable from \eqref{KK_p5_rel_q}.
	
	As in the ordinary case, being the fifth dimension a compact space, we have to assume that $p_5$ is quantized; in order to determine the quantisation law, we need to represent the free particle state in the fifth coordinate as described by the wave function of the polymer free particle. 
	As comprehensively discussed in \cite{Corichi2}, in the position representation, this will be:
	\begin{equation} \label{PG_free_poly_particle_normalized}
		\psi_{\mu}(x^5)=\frac{1}{\sqrt{L}}e^{ ix^5p_{5}^{(\mu)}/\hbar},
	\end{equation}
	normalized on $S^1_{x_5}$, where $L$ is the length of the circus which characterizes the fifth dimension.
	Hence, by imposing $\psi_{\mu}(x^5)$ to be periodic in $x^5$, we achieve the following quantisation law:
	\begin{equation} \label{PG_polymer_momentum_quantisation}
		p_{5}^{(\mu)}= \frac{2\pi n}{L}\hbar  \qquad n \in\mathbb{Z},
	\end{equation}
	which immediately leads, accordingly to \eqref{PG_poly_charge_p5}, to a new quantisation law for the electric charge:
	\begin{equation} \label{PG_charge_quantisation}
		q_{n}\approx\frac{2\hbar\sqrt{G}}{\mu c}\sin(\frac{2\pi \mu}{L}n),
	\end{equation}
	defined in the aforementioned periodic interval.
	
	\noindent We can choose $q_n$ to be equal to the electron electric charge $e$ for $n=1$, as in the ordinary case, and in this way we are able to find an expression for $L$:
	\begin{equation} \label{PG_L_function}
		L\approx\frac{2\pi\mu}{\arcsin(\frac{e\mu c}{2\hbar\sqrt{G}})},
	\end{equation}
	
	from which we deduce a constraint for the polymer scale $\mu$:
	\begin{equation} \label{PG_scale_constraint}
		0 < \mu \leq \frac{2\hbar\sqrt{G}}{\abs{e}c} \approx 3.78 \cdot 10^{-32} \> cm.
	\end{equation}
	At this point we choose $\mu$ equal to the Planck length and obtain: 
	\begin{equation} \label{PG_poly_geo_L}
		L\approx 2.377 \cdot 10^{-31} \> cm,
	\end{equation}
	which almost coincide with the result of the standard theory and therefore it can account for the non observability of the fifth dimension.
	
	There are basically two reasons behind this particular choice of the polymer scale: 
	\begin{itemize}
		\item it is a scale with a strong physical meaning;
		\item the $L/\mu$ ratio, for this value of $\mu$, is large enough to ensure some calculations to be carried out in the polymer continuum limit, through the assumptions discussed in \cite{Corichi1} and \cite{Corichi2}.
	\end{itemize}
	
	\noindent A further discussion about the function \eqref{PG_L_function} is postponed to the next subsection.
	
	\medskip 
	
\begin{figure}[h!]
		\centering
		\includegraphics[width=%
		0.5\textwidth]{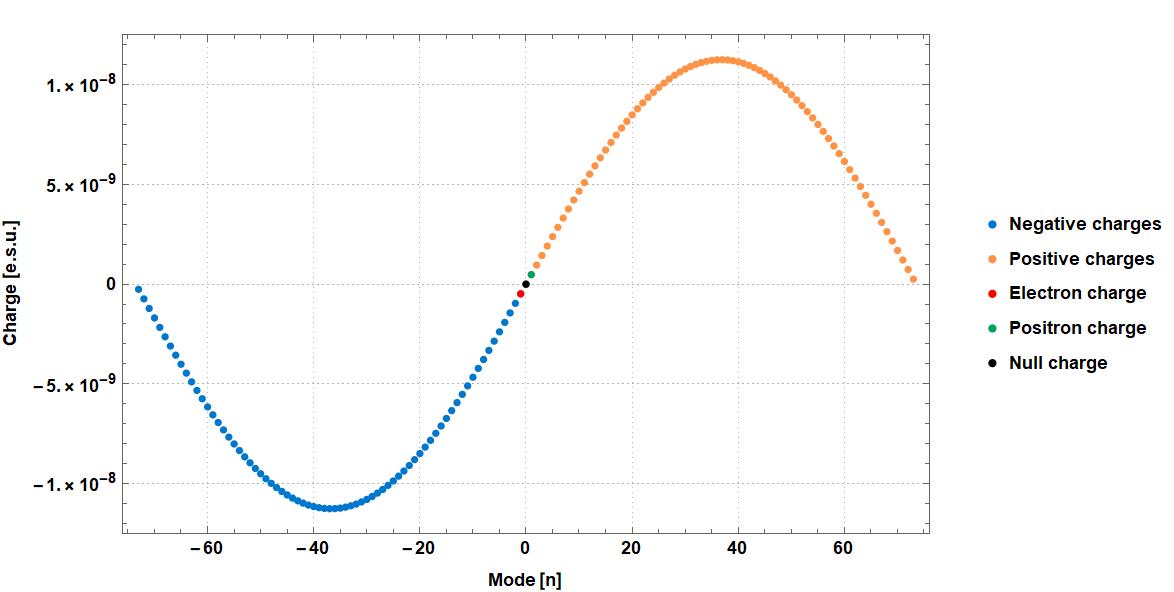}	
		\caption{Plot of charges distribution for a value of $\mu$ equal to the Planck length. It is possible to appreciate the oscillating profile of the function and the negative and positive symmetric branches.}
		\label{PG_planckian_charge_distribution_plt}
	\end{figure}

	The charge function \eqref{PG_poly_charge_p5}, instead, can be rewritten, by means of the function $L(\mu)$ \eqref{PG_L_function}, as follows:
	\begin{equation} \label{PG_charge_distr_func_poly}
		q(\mu;n)=\frac{2\hbar\sqrt{G}}{\mu c}\sin\bigg(n\arcsin(\frac{\mu ec}{2\hbar\sqrt{G}})\bigg)
	\end{equation}
	
	and by setting the polymer scale $\mu$ equal to the Planck length, we obtain the symmetric distribution of positive and negative charges reported in Figure \ref{PG_planckian_charge_distribution_plt}.
	
	\noindent The number of modes $n$ in the considered interval is limited by the periodic condition of the function itself and it clearly depends on $\mu$.
	For our choice of $\mu$ we find that $-73\leq n \leq 73$ (see again Figure \ref{PG_planckian_charge_distribution_plt}). 
	
	\noindent It is worth noticing that for any fixed value of $n=n^*$ it is always possible to expand the sine function in correspondence to a suitable small cut-off parameter $\mu$, accordingly to the inequality $n^*\mu<<2\hbar\sqrt{G}/ec$, where we also have expanded the function $L(\mu)$.
	
	\noindent In this limit we recover the standard expression for the charge as multiple of the elementary electron charge, at least for $n<n*$.

	\medskip

	In the end, we want to evaluate the consequence of the condition $\abs{u_5}<1$ in our framework.  
	
	\noindent The relation between $u_5$ and $q$ can be found employing the expression \eqref{PG_polymer_q_expression}.
	
	\noindent Hence, by retaining only the zero-order term in $u_5$ - coherently with the overall expansion in $A_{\mu}$ and its derivatives we have done in the equation of motion \eqref{PG_4d_polymer_geodesic_equation} - we can write:
	\begin{equation}
		\abs{u_5}\approx \frac{\abs{q}}{2m\sqrt{G}}\sqrt{\frac{1}{2}+\frac{2 G}{q^2m^2}
			\pm \frac{1}{q^2\hbar}F(\mu)},
	\end{equation}
	where:
	\begin{equation}
		F(\mu):=\sqrt{4\mu ^2 m^2c^2 q^4+16G^2\hbar^2m^4-8G \hbar^2 m^2 q^2+h^2 q^4}.
	\end{equation}
	
	\noindent By imposing the condition $\abs{u_5}<1$, the solution with plus sign can be ruled out, since it always violates the constraint, while for the solution with minus sign - which is the meaningful one - we obtain:
	\begin{equation} \label{PG_poly_q/m_constraint}
		\frac{\abs{q}}{m}\leq \frac{2\hbar\sqrt{G}}{\mu m c},
	\end{equation}
	which is in agreement with the quantisation law \eqref{PG_poly_charge_p5}.
	
	\medskip
	
	\noindent This constraint - that we stress is valid only in a perturbative regime - is different from the ordinary one \eqref{KK_q_over_m} and it introduces a dependence in the $q/m$ ratio from the polymer scale $\mu$ - as expected - and from the mass itself.
	Remarkably the empirical $q/m$ ratio of any known particle always respects the bound for every value of $\mu$ in the interval \eqref{PG_scale_constraint}.
	This means that the new constraint does not set an unphysical condition and at least does not contradict the experimental evidence, hence resolving one of the shortcoming aspects of the standard theory. 
	In other words, we are now able, in principle, to reproduce the $q/m$ ratio of any Standard Model particle.

	\subsection{Polymer complex scalar field coupled with Kaluza-Klein metric}
	
	In this subsection we will carry out the study of a complex scalar field polymer dynamics, on a Kaluza-Klein background, obtained through a proper perturbation of the previous polymer Kasner metric.
	
	\medskip 
	
	Following the literature \cite{Chodos-Detweiler}\cite{Montani}\cite{Landau}, under the isotropy assumption in the three usual spatial dimensions, we rewrite the polymer Kasner solution as:
	\begin{equation}
		\begin{split}
			ds^2=&-c^2dt^2+\left(\frac{t}{\tau}\right)^{2k}(d\vec{x})^2+\left(\frac{t}{\tau}\right)^{2k_5}(dx^5)^2,
		\end{split}
	\end{equation}
	where $\tau$ is a time characteristic of the present age of the Universe and we have rescaled the coordinates by the same factor.
	
	\noindent Since we are interested in the static case with respect to the fifth dimension, we choose $k_5=0$ and hence, as we have seen in the previous section, $k=1/3$:
	\begin{equation} \label{PCSF_Kasner_metric}
		ds^2=-c^2dt^2+\left(\frac{t}{\tau}\right)^{2/3}(d\vec{x})^2+(dx^5)^2.
	\end{equation}
	Taken into account the rescaling, the coordinates will now satisfy the following conditions:
	\begin{equation} \label{x5_values}
		0\leq \abs{\vec{x}}<L(\tau/t_0)^{k} \qquad  0\leq x^5<L.
	\end{equation}
	
	\noindent In order to include a perturbative electromagnetic field, a small perturbation $h_{\mu 5}$ ($\mu$=0,1,2,3) is added to the polymer-modified Kasner metric, properly proportional to the electromagnetic field itself:
	\begin{equation}
		h_{\mu 5}=\frac{2\sqrt{G}}{c^2}A_{\mu}.
	\end{equation}
	\noindent As stated in the previous section this is naturally justified in a cosmological setting.
	
	Coupled to this background we introduce a quantum complex scalar field $\Phi$, which ordinary dynamics is described by a $5D$ Klein-Gordon equation and we set the $5D$ "mass" term equal to zero:
	\begin{equation} \label{PCSF_KG_5d_eq}
		^{(5)}\Box \Phi(x^a)=0 \qquad a=0,1,2,3,5.
	\end{equation}
	
	\noindent As in the standard case, our analysis relies on the assumption of the cylinder hypothesis for observable physical quantities, still providing the scalar field $\Phi$ with a phase factor depending on $x^5$, the latter described by the polymer free particle wave function, with periodicity condition on the fifth coordinate due to the topology of the space:
	\begin{equation}
		\Phi(x^{\mu},x^5)=\frac{1}{\sqrt{L}}\phi(x^\mu) e^{ix^52\pi n/L}.
	\end{equation}
	where the function $\phi(x^\mu)$ depends only on the variables of the $4D$ space-time. 
	
	\noindent A 4+1 dimensional splitting of eq.\eqref{PCSF_KG_5d_eq} results in the following expression:
	\begin{equation} \label{PCSF_KG_5d_eq_splitted}
		\begin{split}
			&\partial_\mu\partial^\mu\Phi-\frac{2\sqrt{G}}{c^2}\partial_\mu A^\mu\partial_5\Phi-\frac{4\sqrt{G}}{c^2}A^\mu\partial_\mu\partial_5\Phi\\
			&+\left[1+\frac{4G}{c^4}A_\nu A^\nu\right]\partial_5^2\Phi=0,
		\end{split}
	\end{equation}
	We want now to study eq.\eqref{PCSF_KG_5d_eq_splitted} by introducing polymer quantum framework only with respect to the fifth dimension; therefore we switch to momentum representation only on the fifth coordinate and then let the operator $\hat{p}_5$ act according to the polymer prescription \eqref{PQ_p_op_approx}.
	
	\noindent It is worth noticing that, acting in this way, we are working in a mixed representation of position and momentum, which would not be possible universally speaking, nevertheless, since the coupling term in \eqref{PCSF_KG_5d_eq_splitted} between $x^{\mu}$ and $x^5$ is a small perturbation, we are legitimate, under this assumption, to proceed along this path, according to the diagonal form of the background metric.
	
	\noindent Finally, the following equation for the complex scalar field is achieved:
	\begin{equation} \label{PCSF_poly_KG_field_eq}
		\begin{split}
			&{}^{(4)}\Box\phi(x^\mu)-i\frac{2\sqrt{G}}{c^2\mu}\phi(x^\mu)\partial_{\mu}A^{\mu}\frac{1}{\mu}\sin({\mu}\frac{2\pi n}{L})\\
			&-i\frac{4\sqrt{G}}{c^2\mu}\sin({\mu}\frac{2\pi n}{L})A^{\mu}\partial_\mu\phi(x^\mu)\\
			&-\left[1+\frac{4G}{c^4}A_\nu A^\nu\right]\frac{1}{\mu^2}\sin[2]({\mu}\frac{2\pi n}{L})\phi(x^\mu)=0.
		\end{split}
	\end{equation}
	
	\noindent By comparing term by term the latter with the $4D$ equation of a massive complex scalar field, coupled to an electromagnetic one, on a curved space-time, we attain the following identifications:
	\begin{equation} \label{PCSF_polymer_law_charge_and_mass}
		\begin{split}
			&q_n=\frac{2\hbar \sqrt{G}}{\mu c}\sin(\frac{2\pi \mu}{L}n) \\
			&m_n=\frac{\hbar}{\mu c}\abs{\sin(\frac{2\pi \mu}{L}n)}.
		\end{split}
	\end{equation}
	\noindent The first one is again a quantisation law for the electric charge and it is clearly the same we have obtained in \eqref{PG_charge_quantisation}, in the modified geodesic study.
	
	\noindent Hence, by imposing once again $q_n=e$ for $n=1$, we recover for the size $L$ of the fifth dimension the function \eqref{PG_L_function}, with the constraint \eqref{PG_scale_constraint}.
	
\begin{figure}[h!]
		\centering
		\includegraphics[width=%
		0.5\textwidth]{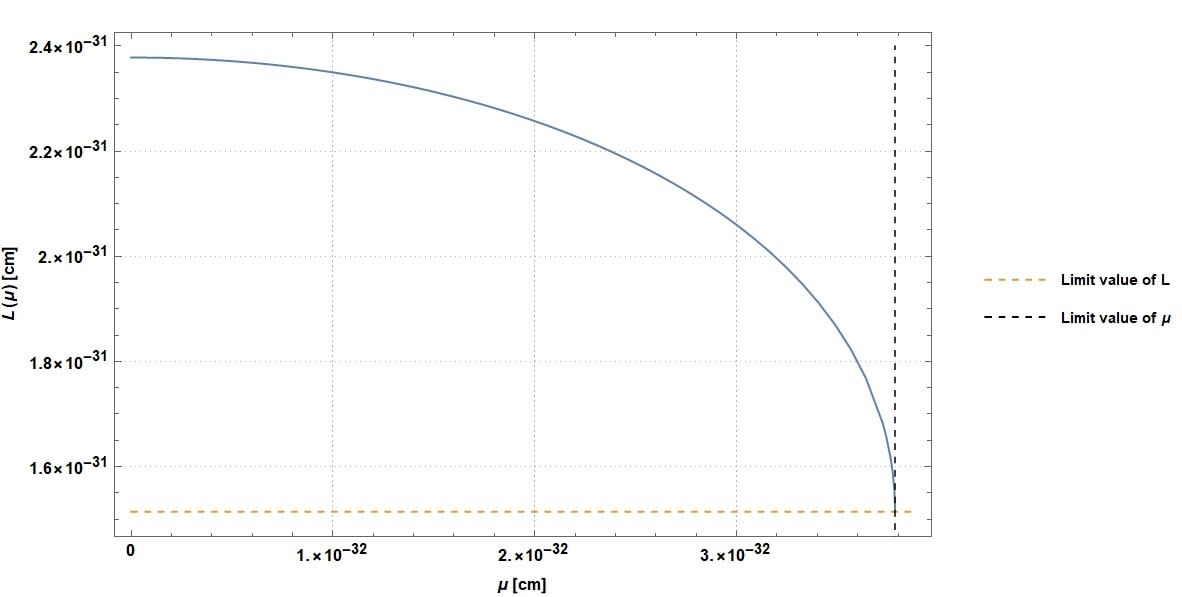}	
		\caption{Monotonically decreasing trend of the size $L$ of the fifth dimension as a function of the polymer scale $\mu$. The constraint \eqref{PG_scale_constraint} for $\mu$ establishes a precise domain and codomain for the function $L(\mu)$.}
		\label{L_function_plt}
	\end{figure}

	\noindent This function is a monotonically decreasing function of $\mu$, which reaches its minimum value at the right endpoint of the interval in which is defined and it tends exactly to the value of the ordinary theory \eqref{KK_ordinary_L} as $\mu$ approaches zero (see Figure \ref{L_function_plt}).
	
	\noindent Clearly, choosing again $\mu$ around the Planck length, we find for the value of $L$ the result \eqref {PG_poly_geo_L}, discussed above.
	
	\noindent On the other hand, the second relation \eqref{PCSF_polymer_law_charge_and_mass} represents a mass distribution law for the scalar field, which, by means of the expression \eqref{PG_L_function}, can be rewritten as a function
	of the integer parameter $n$ and of the continuous variable $\mu$:
	\begin{equation} \label{mass_distr_func_poly}
		m(\mu;n)=\frac{\hbar}{\mu c}\abs{\sin\bigg(n\arcsin(\frac{\mu ec}{2\hbar\sqrt{G}})\bigg)}.
	\end{equation}
	
	\noindent This function - for the same reasons stated in the previous subsection regarding the charge function - can be defined in the periodic interval $[-\pi \hbar/\mu,\pi \hbar/\mu]$, but, because of the presence of the modulus, we can restrict our attention only on the positive segment of the interval itself, that is $[0, \pi \hbar/\mu]$.
	For a fixed value of the polymer scale $\mu$, the resulting sequence of masses, due to the periodicity condition, will include only a finite number of mass modes $n$ and it clearly will result bounded and oscillating, as $n$ changes. 
	
	\noindent This freedom in the choice of the scale leads to a crucial feature of such distribution: picked up a particular mode $n$, through a fine-tuning of $\mu$, it is possible to fit any desired value of the mass, therefore also Standard Model-comparable masses. 
	In particular, the spectrum can be explored in correspondence to a Planckian-like value of the parameter $\mu$. 
	
\begin{figure}[h!]
		\centering
		\includegraphics[width=%
		0.5\textwidth]{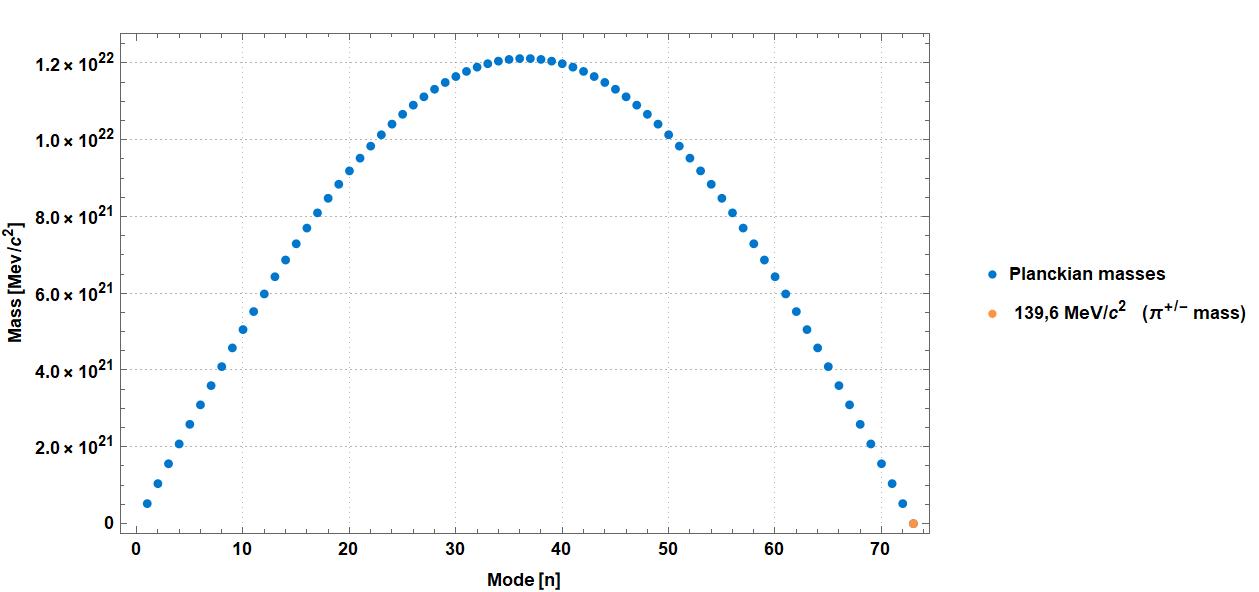}	
		\caption{Plot of masses distribution for the complex scalar field, for the value $\mu_{Planck}\approx 1.627\cdot10^{-33} \> cm$. It is possible to observe the oscillating profile of the function and the fitted Pion mass, placed in the minimum, as ground level of the Kaluza-Klein tower.}
		\label{planckian_masses_distribution_plt}
	\end{figure}

	\noindent The specific pattern of the distribution then will depend on the chosen reference mass to be fitted to a certain mode $n$, for a fixed value of $\mu$; nevertheless, it can be shown that a qualitative general trend can be recovered, which provides the presence of the assigned mass in the minimum of the sequence as a ground level of the Kaluza-Klein tower.
	We stress that these regular values appear accompanied by Planckian masses, as sketched in Figure \ref{planckian_masses_distribution_plt}.
	
	\noindent This distribution deeply differs from the one obtained in the standard theory, reported in \cite{Chodos-Detweiler}.
	In fact, the latter is a linear and increasing sequence of only Planckian masses, without an upper limit.
	
	\noindent Actually, the possibility of fitting arbitrary masses and from this obtaining various distributions can be achieved also in the ordinary theory, as shown in \cite{Chodos-Detweiler}, by introducing a $5D$ mass term $a$ in the Klein-Gordon equation \eqref{PCSF_KG_5d_eq}, which role is that of an additive parameter to be fine-tuned.
	
	\noindent Nevertheless, this way of proceeding reveals several issues, which consequently are all addressed in our framework, not being necessary the introduction of such an extra quantity $a$.
	
	\noindent First of all, in general, the fine-tuning procedure cannot be applied to every mode, since the linear and increasing trend of the distribution is not modified at all; in the second place, the addition of this parameter eventually generates the rise of tachyonic masses in the past.
	This happens because of the presence of the metric coefficient $(t/\tau)^{\abs{k_5}}$ in the standard momentum expression, as discussed in \cite{Chodos-Detweiler}, which cannot be removed, since in the ordinary theory the Kasner solution does not admit the case $k_5=0$. 
	
	\noindent This means that in our framework the tachyonic masses are ruled out from the polymer-modified distribution. 
	
	\smallskip
	
	We outline that - in our analysis - even in the case $k_5\neq 0$, i.e. in the non-static case, the tachyonic problem would be solved since ultimately it is the chance to set $a=0$ which removes the unphysical masses.
	
	\smallskip
	
	Finally, the $a$ parameter has to be introduced on purpose in the theory and its physical meaning remains ambiguous, while in our framework the fine-tuning parameter is an internal degree of freedom of the theory.
	
	\medskip
	
	Hence, even if polymer quantum mechanics fails in removing completely the Planckian masses from the spectrum, it succeeds first in introducing a cut-off since the masses distribution ref is bounded.
	Furthermore, in the considered natural interval a finite number of variable mass values take place. 
	in the masses distribution in the considered periodic interval of definition of the masses function itself and secondly in accounting for Standard Model-comparable masses, under a proper fine-tuned choice of the scale variable $\mu$.
	
	\medskip 
	
	Since we are dealing with a complex scalar field, the charged pions $\pi^{\pm}$ seem to be relevant (phenomenologically) candidates in Nature for the reference masses.
	
	\noindent It results that the pion mass can be fitted at the mode $n=73$ for a fine-tuned value of $\mu$ (up to twenty decimal figures) almost equal to the Planck length, which we will denote from now on as $\mu_{Planck}$.
	
	\noindent Nevertheless, for this procedure to be consistent, it is necessary to verify that the resulting mode $n=73$ belongs to the range of $n$ admitted in the periodic interval of definition of the masses function.
	
	\noindent It is easy to show that for $\mu=\mu_{Planck}$ the condition for positive $n$ for belonging - in which we are interested - is $n\leq73$.
	
	\noindent Therefore, the results coming from the fit procedure are valid and legitimate in this regard.
	
	In Figure \ref{planckian_masses_distribution_plt} it is reported the whole sequence of masses resulting for $\mu=\mu_{Planck}$, where it is possible to appreciate the general behaviour discussed above.
	
	\noindent In particular, we observe that the pion mass is in the minimum of the distribution and that the maximum available mass is always: \[m_{max}=\frac{\hbar}{\mu c}\]
	which is inversely proportional to $\mu$, while the first mode of the sequence ($n=1$) is: 
	\begin{equation}
		m(\mu;1)=\frac{e}{2\sqrt{G}}\approx 9.29\cdot 10^{-7} \> g.
	\end{equation}
	
	\noindent It does not depend on $\mu$ and it is almost a Planckian mass, defined only by fundamental constants, equal to the one obtained in the standard framework (without the introduction of the \textit{ad hoc} parameter $a$).
	
\begin{figure}[h!]
		\centering
		\includegraphics[width=%
		0.5\textwidth]{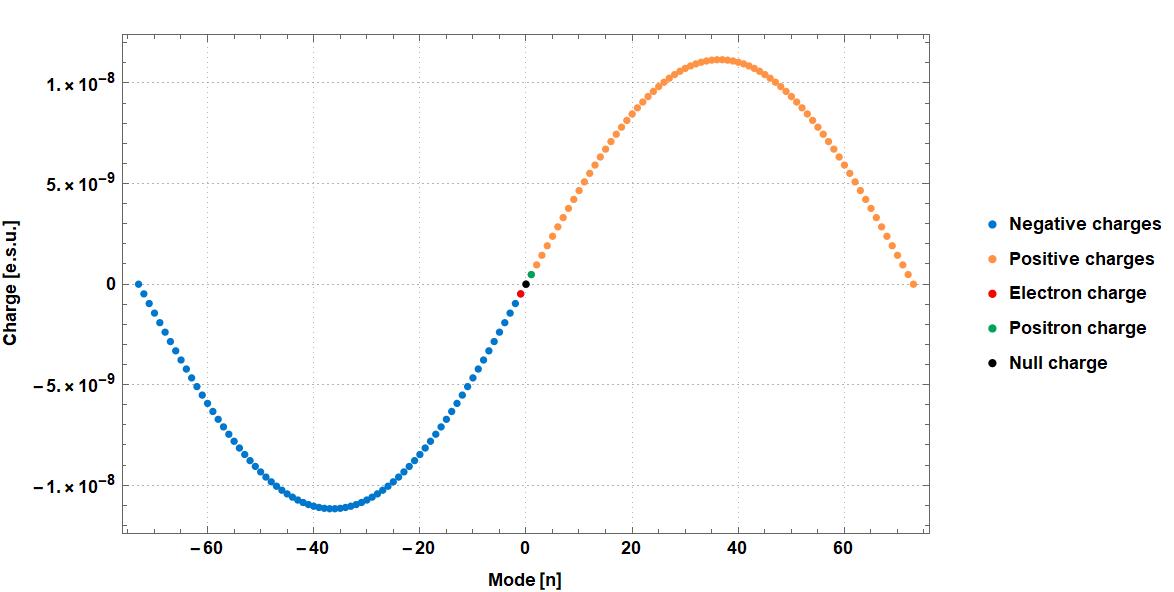}	
		\caption{Plot of charges distribution for value $\mu=\mu_{Planck}$. Again, it is possible to observe the oscillating profile of the sequence and the negative and positive symmetric branches.}
		\label{KG_planckian_charge_distribution_plt}
	\end{figure}

	In Figure \ref{KG_planckian_charge_distribution_plt} instead the corresponding charge distribution - which again can be put in the form \eqref{PG_charge_distr_func_poly} - for $\mu=\mu_{Planck}$ is represented.

	\noindent We observe, however, that only the electron charge ($n=1$) has a phenomenological correspondence, while the remaining points of the sequence have not a clear interpretation.
	In particular according to these mass and charge distributions the fundamental charge $e$ has to be associated with a particle of Planckian mass, while the corresponding charge of the pion would be several orders of magnitude smaller than the electron charge.

	\noindent Clearly, this peculiar charge-mass configuration is not phenomenological consistent.
	
	\noindent Indeed, calculating the $q/m$ ratio for the modes of the scalar field, we obtain:
	\begin{equation}
		\frac{q(\mu;n)}{m(\mu;n)}=\pm 2\sqrt{G}\approx 5.16\cdot 10^{-4} e.s.u./g,
	\end{equation}
	where the $\pm$ sign is due to the sign of the electric charge.
	
	\noindent This value, which coincides with the upper limit of the $q/m$ ratio \eqref{KK_q_over_m} of the standard classical case, does not depend on $\mu$ nor $n$, rather it is constant for every polymer scale and every mode and no known particle satisfies such a relation.

	\section{Conclusion}
	\label{Conlusion}
	
	We investigated the formulation of a five-dimensional Kaluza-Klein theory in the framework of Polymer Quantum Mechanics, viewed both in a semi-classical and quantum approach. 
	The polymer modifications have been implemented to the fifth coordinate only, on a semi-classical level in the spirit of the Ehrenfest theorem (the modification provides the dynamics of the quantum expectation values) and 
	in a full quantum approach when a Klein-Gordon equation has been investigated. 
	
	\medskip 
	
	We started by applying the semi-classical polymer formulation to the evolution of the Bianchi I model, by showing that the corresponding Kasner solution can be taken in a form in which three scale factors isotropically expand and the remaining one is static and, in the considered model, it coincides with the compactified extra-dimension. 
	
	\noindent Then we studied the geodesic motion of a particle, starting with a Hamiltonian formulation (the only one in which the polymer formulation is viable) and then turning to a formalism based on the particle velocities.
	This procedure allows, in analogy to the standard literature on this same subject, to identify the expression for the electric charge via the fifth momentum component of the particle. 
	The important consequence of this revised formulation consists of the overcoming of the problem of a too small charge to mass ratio to account in the model for any known elementary particle. 
	In fact, the revised constraint, due to the polymer relation between the fifth momentum component and the corresponding velocity, is in principle compatible with all the elementary particle predicted by the Standard Model. 
	
	\noindent Finally, we implemented a quantum polymer modification in the Klein-Gordon equation, by adopting a mixed representation of quantum mechanics (based on the coordinates for the usual four dimensions and the momentum for the extra one). 
	This study has the aim to revise the analysis in \cite{Chodos-Detweiler} for a static (now available) extra-dimension, under a polymer prescription for the 
	compactified dimension physics. 
	
	We got the fundamental result that the tachyon mode, present in \cite{Chodos-Detweiler} is now removed from the mass spectrum and that the obtained values for the boson mass can fit the values spanned in the Standard Model. 
	Actually, we arrived at a deformed morphology of the so-called Kaluza-Klein tower (the steps are no longer equispaced), but this revised structure allows us to avoid the only Planckian mode naturally present in the standard Kaluza-Klein formulation. 
	
	\medskip 
	
	All these results suggest that some of the puzzling question affecting the viability of the Kaluza-Klein idea must be reanalysed phenomenologically including the notion of a cut-off physics. 
	In fact, in the case of small dimensions, living about two orders of magnitude over the Planck scale, it should be unavoidable to feel the effects of the nearby cut-off and when its presence is made manifest a new paradigm can be assessed.
	In other words, we argue that some limits of the geometrical unification theories are possibly due to the ultraviolet divergence that the gravitational field possesses and when they are somehow attenuated, like by the polymer scenario adopted here, the compactified dimension takes a more regular behaviour, which is reflected into the solution of some inconsistencies of the underlying model.
	
	The emergence of a static dimension in the $5D$ Kasner solution - which prevents the necessity to deal with unphysical tachyonic modes - undoubtedly represents the simplest elucidation of this point of view.

	\bibliography{GMSS_article}
\end{document}